\documentclass[runningheads]{llncs}
\usepackage[T1]{fontenc}
\usepackage{graphicx}
\newcommand{\AltTextCMSB}[1]{}

\usepackage[T1]{fontenc}
\usepackage{graphicx}

\usepackage{lineno}

\usepackage{hyperref}
\usepackage{url}
\usepackage{xcolor}
\usepackage{mathtools}
\usepackage{bussproofs}

\usepackage{xcolor}
\usepackage{caption}

\usepackage{listings}

\lstdefinelanguage{ASP}{
    morekeywords={:-,not},
    sensitive=true,
    morecomment=[l]\%,
    morestring=[b]",
}

\lstdefinestyle{myasp}{
    language=ASP,
    basicstyle=\ttfamily\small,
    keywordstyle=\color{blue},
    commentstyle=\color{gray},
    stringstyle=\color{red},
    showstringspaces=false,
    breaklines=true,
    frame=single,
    captionpos=b
}

\usepackage{microtype}
\microtypesetup{final,tracking=true,kerning=true,spacing=true,stretch=10,shrink=10,letterspace=25}

\usepackage[english]{babel}
\newcommand{\mathdefault}[1][]{}

\newcommand{\ignore}[1]{}

\newif\ifcleanbuild
\cleanbuildtrue %

\ifcleanbuild

\newcommand{\newMat}[1]{}
\newcommand{\todoR}[1]{}
\newcommand{\todo}[1]{}
\newcommand{\todoH}[1]{}
\newcommand{\todoA}[1]{}
\newcommand{\todoN}[1]{}
\newcommand{\nanis}[1]{}
\newcommand{\nanisH}[1]{}
\newcommand{\hans}[1]{}
\newcommand{\hansN}[1]{}
\newcommand{\alt}[1]{}

\else

\newcommand{\newMat}[1]{\color{green}{#1}\color{black}}
\newcommand{\todoR}[1]{\color{red}{#1}\color{black}}
\newcommand{\todo}[1]{\textcolor{red}{#1}}
\newcommand{\todoH}[1]{\textcolor{violet}{TODOH : #1}}
\newcommand{\todoA}[1]{\textcolor{olive}{TODOA : #1}}
\newcommand{\todoN}[1]{\textcolor{olive}{TODOA : #1}}
\newcommand{\nanis}[1]{\textcolor{olive}{#1}}
\newcommand{\nanisH}[1]{\textcolor{olive}{@Hans: #1}}
\newcommand{\hans}[1]{\textcolor{violet}{#1}}
\newcommand{\hansN}[1]{\textcolor{violet}{@Athénaïs: #1}}
\newcommand{\alt}[1]{\textcolor{orange}{ALTERNATIVE: #1}}
\linenumbers

\fi

\newcommand{\todoI}[1]{}

\newcommand{\todoFINAL}[1]{}

\newcommand{\classic}[1]{\textsf{classic}}
\newcommand{\ournoMA}[1]{\textsf{our noMA}}
\newcommand{\ourMA}[1]{\textsf{our MA}}

\usepackage{biblatex}
\AtEveryCitekey{\defcounter{maxnames}{1}}

\AtBeginDocument{
    \AtEveryBibitem{\clearfield{url}}
    \AtEveryBibitem{\clearfield{urldate}}
    \AtEveryBibitem{\clearfield{urlyear}\clearfield{urlmonth}\clearfield{urlday}}
    \AtEveryBibitem{\clearfield{abstract}}
    \AtEveryBibitem{\clearfield{langid}}
    \AtEveryBibitem{\clearfield{keywords}}
    \AtEveryBibitem{\clearfield{file}}
    \AtEveryBibitem{\clearfield{month}}
    \AtEveryBibitem{\clearfield{day}}
    \AtEveryBibitem{\clearfield{pmcid}}
    \AtEveryBibitem{\clearfield{isbn}}
    \AtEveryBibitem{\clearfield{issn}}

    \AtEveryCitekey{\clearlist{location}}
    \AtEveryBibitem{\clearlist{location}}
}

\usepackage{subcaption}

\usepackage{tikz}
\usetikzlibrary{arrows}

\usepackage{xspace}

\usepackage{amsmath}
\usepackage{amssymb}

\usepackage[capitalise]{cleveref} %

\newcommand{\setreal}{\ensuremath\mathbb{R}}
\newcommand{\setreals}{\setreal}
\newcommand{\setR}{\setreal}

\newcommand{\setsign}{\ensuremath\mathbb{S}}
\newcommand{\setS}{\setsign}

\newcommand{\setRp}{\setreal_+}

\DeclareMathOperator{\signUP}{\uparrow}
\DeclareMathOperator{\signDO}{\downarrow}
\DeclareMathOperator{\signNO}{0}

\DeclareMathOperator{\sign}{\mathsf{sign}}

\newcommand{\T}{\boldsymbol{1}}
\newcommand{\F}{\boldsymbol{0}}

\newcommand{\varspecies}[1]{\ensuremath\mathsf{#1}}

\newcommand{\speciesA}{\ensuremath\mathsf{A}}
\newcommand{\speciesB}{\ensuremath\mathsf{B}}
\newcommand{\speciesC}{\ensuremath\mathsf{C}}

\newcommand{\speciesE}{\ensuremath\mathsf{E}}
\newcommand{\speciesES}{\ensuremath\mathsf{C}}
\newcommand{\speciesS}{\ensuremath\mathsf{S}}
\newcommand{\speciesP}{\ensuremath\mathsf{P}}
\newcommand{\speciesX}{\ensuremath\mathsf{X}}
\newcommand{\speciesY}{\ensuremath\mathsf{Y}}

\newcommand{\setSpecies}{\ensuremath{\mathcal{S}}}
\newcommand{\setspecies}{\setSpecies}
\newcommand{\setspec}{\setSpecies}

\newcommand{\setDSpecies}{\ensuremath{\dot{\mathcal{S}}}}

\newcommand{\setDSpec}{\setDSpecies}

\newcommand{\setNSpecies}{\ensuremath{\next{\mathcal{S}}}}
\newcommand{\setNspecies}{\setNSpecies}

\newcommand{\setDNSpecies}{\ensuremath{\next{\dot{\mathcal{S}}}}}

\newcommand{\setDNSpec}{\setDNSpecies}

\newcommand{\SigmaArith}{\ensuremath\Sigma_{\mathrm{arith}}}

\newcommand{\arithmOps}{\ensuremath{\{+, -, *,/\}}}

\newcommand{\reaction}{\ensuremath{\mathcal{R}}}
\newcommand{\reac}{\reaction}
\newcommand{\react}{\reaction}

\newcommand{\reactnet}{\ensuremath{\boldsymbol{\mathcal{R}}}}
\newcommand{\rn}{\reactnet}

\newcommand{\fobnn}{\ensuremath{\mathcal{F}}}
\newcommand{\pbnn}{\ensuremath{\mathcal{P}}}

\newcommand{\phiRenz}{\fobnn_{\Renz}}

\DeclareMathOperator{\tto}{\ensuremath{\text{\tikz[baseline=-0.5ex]\draw[-latex] (0,0) -- +(0.35,0);}}}

\usepackage{wasysym}

\newcommand{\odes}{\ensuremath\mathsf{ode}}
\newcommand{\odesR}{\ensuremath{\odes_{\rn}}}
\newcommand{\odeR}{{\odesR}}

\newcommand{\config}{\sigma}
\newcommand{\state}{\config}

\newcommand{\bigland}{\ensuremath\bigwedge}

\renewcommand{\next}[1]{\ensuremath {#1}_{\mathrm{next}}}
\newcommand{\nextdot}[1]{\next{\dot{#1}}}
\newcommand{\dotnext}[1]{\next{\dot{#1}}}

\newcommand{\nextX}{\next{\speciesX}}
\newcommand{\dotX}{\dot{\speciesX}}

\newcommand{\eqdef}{:=}

\newcommand{\eq}{\ensuremath{\eqcirc}} %
\newcommand{\EQ}{\ensuremath{\eqcirc}} %

\newcommand{\limp}{\Rightarrow}

\newcommand{\setVars}{\ensuremath\mathcal{V}}

\newcommand{\setTerms}{\ensuremath\mathcal{T}}

\newcommand{\setTermsArith}{\ensuremath\setTerms_{\mathrm{arith}}}

\newcommand{\formu}{\ensuremath\varphi}
\newcommand{\calI}{\ensuremath\mathcal{I}}

\newcommand{\sem}[2]{[\! [#1] \!]_{#2}}

\newcommand{\restrict}[2]{\ensuremath{#1}_{|#2}}

\newcommand{\restr}[2]{\restrict{#1}{#2}}

\newcommand{\FP}[2]{\ensuremath{\mathrm{FP}}}

\newcommand{\Renz}{\reactnet_{\mathrm{enz}}}

\newcommand{\varkon}{\ensuremath{\mathrm{k}}_{\mathrm{on}}}
\newcommand{\varkoff}{\ensuremath{\mathrm{k}}_{\mathrm{off}}}
\newcommand{\varkcat}{\ensuremath{\mathrm{k}}_{\mathrm{cat}}}

\newcommand{\eon}{\ensuremath{ \varkon * {\varspecies{S}} * {\varspecies{E}} }}
\newcommand{\eoff}{\ensuremath{\varkoff * {\varspecies{C}}  }}
\newcommand{\ecat}{\ensuremath{ \varkcat* {\varspecies{C}} }}

\newcommand{\nexteon}{\ensuremath{ \varkon * {\next{\varspecies{S}}} * {\next{\varspecies{E}}} }}
\newcommand{\nexteoff}{\ensuremath{\varkoff * {\next{\varspecies{C}}}  }}
\newcommand{\nextecat}{\ensuremath{ \varkcat* {\next{\varspecies{C}}} }}

\newcommand{\ron}{\ensuremath{\react_{\mathrm{on}}}}
\newcommand{\roff}{\ensuremath{\react_{\mathrm{off}}}}
\newcommand{\rcat}{\ensuremath{\react_{\mathrm{cat}}}}

\begin{document}

\title{Looking for Signs:\\Reasoning About FOBNNs Using SAT}
\titlerunning{Reasoning About FOBNNs Using SAT}

\author{
    Hans-Jörg Schurr\inst{2}\orcidID{0000-0002-0829-5056}
    \\ \and
    Athénaïs Vaginay\inst{1}\orcidID{0000-0001-5062-7993}
}
\authorrunning{H.-J.\ Schurr and A.\ Vaginay}
\institute{
    The University of Iowa\\
    \email{hansjoerg-schurr@uiowa.edu}
    \and
    Université Caen Normandie, ENSICAEN, CNRS, Normandie Univ, 
    \\ GREYC UMR 6072, F-14000 Caen, France\\
    \email{athenais.vaginay@unicaen.fr}
}
\maketitle

\todoI{
    The original FO-BNN paper : \url{https://hal.science/hal-02279942v9/document}
    To read (not sure, but might be interesting)
    \url{https://drops.dagstuhl.de/opus/volltexte/2021/14019/pdf/OASIcs-AUTOMATA-2021-10.pdf}
    Other title ideas :
    \begin{itemize}
        \item FOBNNs as propositional finite state systems
    \end{itemize}
}

\begin{abstract}
    First-Order Boolean Networks with Non-deterministic updates (FOBNN)
compute a boolean transition graph representing the absence and presence of
species over time.
The utility of FOBNNs has been justified by their theoretical soundness
with respect to the Euler simulation of the differential equations.
However,
we lack practical means to work with FOBNNs
and an empirical evaluation of their properties.
We present
a sound and efficient reduction
of the first-order FOBNN transition relation
to a propositional logic formula.
This makes it possible to use modern SAT solvers to reason on the full transition graph, even for large models.
We use this encoding to assess the feasibility and efficiency of
practical reasoning with FOBNNs.
To do so, we focus on the computation of fixed points.
We also compare the transition graphs obtained via FOBNNs
to those computed by the classic boolean semantics of reaction networks.
Overall, our encoding opens new directions for the analysis of FOBNNs and deepens
the understanding of their relationship with reaction networks.

    \keywords{
        First-Order Boolean Networks
        \and
        SAT Solving
        \and
        Reaction Networks
        \and
        Systems Biology
        \and
        Abstract Interpretation
    }
\end{abstract}

\todoI{
    \begin{itemize}
        \item fixed point (instead of fixpoint)
        \item dynamical system (instead of dynamic system)
    \end{itemize}
}

\section{Introduction}

Many formalisms to model biological systems have been proposed.
Boolean abstraction is widely used for reasoning qualitatively
about the dynamics of biological system~\cite{pauleve_reconciling_2020,fagesAbstractInterpretationTypes2008}.

\begin{figure}[t]
    \includegraphics[width=\textwidth]{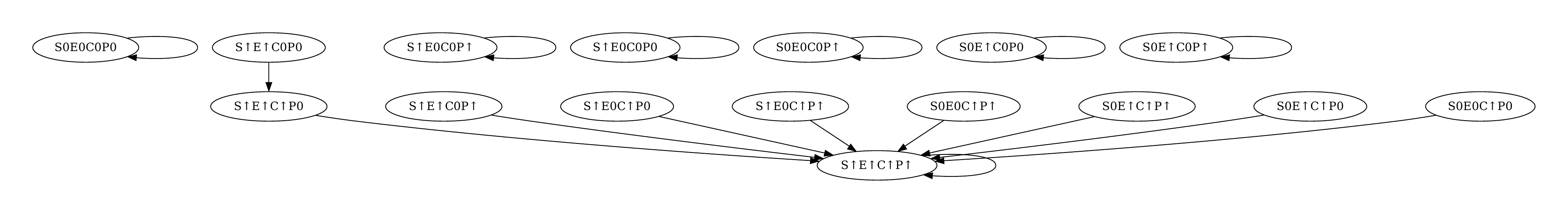}
    \caption{
    \label{fig-Renz-fobnn-stg-withMA}
    \label{Renz-stg-MA}
    $\Renz$ FOBNN transitions
    with mass action constraints (see \cref{sec:additional-constraints:kinetics}).
    The values $\signUP$ and $\signNO$ respectively mean a species is present or absent.
    \AltTextCMSB{
        State transition graph of $\Renz$ on species variables.
    }
    }
\end{figure}

In this paper, we explore
First-Order Boolean Networks with Non-deterministic updates (FOBNNs),
a recently proposed formalism for reaction networks~\cite{niehrenAbstractSimulationReaction2022}. 
FOBNNs use first-order logic to encode a transition relation on
variables representing the real values of species,
and their derivatives~\cite{niehrenAbstractSimulationReaction2022}.
To reason about these first-order interpretation of these formulas
efficiently, the actual values of the parameters and variables are abstracted,
and uses only their signs are used.
Ultimately, an FOBNN computes a boolean transition graph,
such as the one shown in \Cref{Renz-stg-MA}.
However, FOBNN differ from both boolean (automata) networks,
which model with the influences between the species using boolean functions~\cite{schwab_concepts_2020},
and from the classic boolean semantics of reaction networks, which abstract the
discrete semantics of reaction networks~\cite{fagesAbstractInterpretationTypes2008,calzone_biocham_2006}.

FOBNNs can be constructed from the ODEs corresponding to a reaction network.
The resulting transition graph is a sound over-approximation of the traces
produced by the Euler integration of the original ODEs.
Additional first-order constraints can be used to refine the abstraction.
These constraints are either entailed by the full interpretation
of the FOBNN, or encode domain knowledge. 
While this is a strong theoretical justification of the utility of FOBNNs,
we lack practical means to work with FOBNNs,
and an empirical evaluation.

To address this we use a sound and complete reduction of
the first-order FOBNN transition relation to a propositional logic formula whose
models correspond to the models of the original FOBNN formula.
This allows us to harness the power of modern SAT solvers to study the dynamics of FOBNNs.
For example, we can enumerate the transitions of the FOBNN
by enumerating the models of the resulting propositional formula.
Propositional representations of transition systems are also the target domain of SAT-based \emph{model checking}~\cite{biere2018}.
Hence, our work is the first step towards applying these methods to FOBNNs.
Using the encoding, we perform an experimental evaluation of FOBNNs on models from the Biomodels repository.
To do so, we first assess the feasibility and efficiency of our SAT-based approach
to analyze the dynamics of reaction networks.
We focus on the computation of fixed points 
(self-looping states
from which the system cannot escape, e.g.,
the state where all the species are present, 
noted $[\speciesS=\signUP, \speciesE=\signUP, \speciesC=\signUP, \speciesP=\signUP]$ in \Cref{Renz-stg-MA})
as it is one of the most standard inspection of dynamical system~%
\cite{dubrovaSATBasedAlgorithmFinding2011,huDetectionPreventionDeadlock2011,terjelinesEfficientComputationSteady2019}.
Indeed, fixed points in cognate formalisms correspond to
the dynamics of the underlying biological system
(observable biological phenotypes~\cite{remyModelingApproachExplain2015},
observable phase in the cell cycle~\cite{liYeastCellcycleNetwork2004},
or different cell types after differentiation, for example).
They are also often
crucial for applications such as the control of the system~\cite{suCABEANSoftwareControl2021}.

We also compare the transition graphs obtained from FOBNNs
to those computed by the classic boolean semantics of reaction networks.
In particular, we measure the density of the their transition graphs, which
determine how feasible it is to derived meaningful conclusion from the
graph.
In the case of FOBNNs the density can be reduced by encoding additional domain
knowledge.

Our results demonstrate that FOBNNs offer a scaleable and meaningful over-approximation of ODE dynamics,
and provide a new tool to connect continuous and discrete models of biochemical systems.

\paragraph{Outline.}
The paper is structured as follows.
In \cref{sec:prelim}, we present first-order boolean network with non-deterministic updates (FOBNN).
We define their syntax (based on first-order logics) and semantics.
\Cref{sec:encoding} presents our translation of FOBNN into propositional logic.
\Cref{sec:reasoning} describes how the formula is used for practical tasks, such as enumerating transitions and computing fixed points.
In \cref{sec:evaluation} we evaluate the practicality of our method,
and compare its state transition graph to those computed by the classic boolean
semantics of reaction network.
We conclude in \cref{sec:ccl} with a discussion of future directions.



\section{Preliminaries}
\label{sec:prelim}

In this section,
we introduce the concepts needed for the rest of the paper.
In particular, we present
the signature
$\SigmaArith$
and
first-order logic with equality over
this signature. 
This is used to define reaction networks,
differential equations
and FOBNNs.

\subsection{Arithmetic Signature, Terms, Atoms and Formulas}
\label{sec-arith-terms}

The arithmetic signature $\SigmaArith$ is a set $\setR \cup \arithmOps$,
where
$\setR$
is the set of constant symbol and \arithmOps{} a set of function symbols.

The set of arithmetic terms  $\setTerms_{\SigmaArith, \setVars}$
of arithmetic terms over the signature $\SigmaArith$
and variables $\setVars$
  is built using the production rule
\[
  t ::=
  v \mid
  c \mid
  f(t_1, t_2)
  \text{ where }
  v \in \setVars,
  c \in \setR,
  f \in \arithmOps,
  t_1, t_2 \in \setTerms_{\SigmaArith, \setVars}.
\]

\noindent
We will use the common infix notation and precedence for arithmetic expressions,
and we write $-t$ for $0 - t$.

An \emph{atom} $t_1 \EQ t_2$ is the \emph{equality} symbol applied to two terms
$t_1$ and $t_2$.
First-order logic \emph{formulas} $\phi$ are built from atoms:
\[
\phi ::=
    t_1 \EQ t_2 \mid
    \lnot \phi \mid
    \phi_1 \land \phi_2 \mid
    \phi_1 \lor \phi_2 \mid
    \forall x. \phi \mid
    \exists x. \phi
\]

\noindent
We will use the shorthand
$t \geq 0$ for the formula $\exists x. t \EQ x * x $.

The \emph{semantics} 
of first-order terms and formulas is determined by an \emph{interpretation} $\calI$
and a variable \emph{assignment} $\alpha$ on a universe $U$.
The assignment $\alpha$ maps each variable $v$ to $\alpha(v)\in U$.
Similarly, the interpretation $\calI$ maps each
constant $c$ to a value $\calI(c) \in U$, and
each function symbols to a relation\footnote{It is more common to interpret function symbols as functions, but relations are convenient in this paper. Note that we could construct a functional interpretation using the set $2^U$.} on $U$.

Based on $\calI$ and $\alpha$ we can define the interpretation function $\sem{.}{\calI, \alpha}$.
It maps constants to singleton sets (i.e., $\sem{c}{\calI, \alpha} := \{\calI(c)\}$)
and 
operators $\circ\in\Sigma_{\mathrm{arith}}$ to relations:
$
  \sem{t_1 \circ t_2}{\calI, \alpha}
  := \{   s \mid 
    s_1 \in \sem{t_1}{\calI, \alpha},
    s_2 \in \sem{t_2}{\calI, \alpha}, 
    (s_1, s_2, s) \in \calI(\circ)
  \}
$.
This interpretation function for terms can be extended to interpret formulas.
The truth value of an atomic formula is
$\sem{t_1 \EQ t_2}{\calI, \alpha} := \T$ if
          $\sem{t_1}{\calI, \alpha} \cap \sem{t_2}{\calI, \alpha} \neq \emptyset$,
          else $\sem{t_1 \EQ t_2}{\calI, \alpha} := \F$.
Finally, the interpretation of non-atomic formulas is as usual:
\begin{itemize}
    \item $\sem{\exists v.\, \formu}{\calI, \alpha} := \T$
      if there is $s \in D$ with $\sem{\formu}{\calI, \alpha[v \to s]} = \T$,
      else $\sem{\exists v.\, \formu}{\calI, \alpha} := \F$,
    \item $\sem{\forall v.\, \formu}{\calI, \alpha} := \T$
      if for all $s \in D$,  $\sem{\formu}{\calI, \alpha[v \to s]} = \T$,
      else $\sem{\forall v.\, \formu}{\calI, \alpha} := \F$,
  \item $\sem{\neg\formu}{\calI, \alpha} := \T$ if $\sem{\formu}{\calI, \alpha} = \F$,
    else $\sem{\neg\formu}{\calI, \alpha} := \F$,
  \item $\sem{\formu_1 \land \formu_2}{\calI, \alpha} := \T$
    if $\sem{\formu_1}{\calI, \alpha} = \T$ and $\sem{\formu_2}{\calI, \alpha} = \T$,
    else $\sem{\formu_1 \land \formu_2}{\calI, \alpha} := \F$,
  \item $\sem{\formu_1 \lor \formu_2}{\calI, \alpha} := \T$
        if $\sem{\formu_1}{\calI, \alpha} = \T$ or $\sem{\formu_2}{\calI, \alpha} = \T$,
       else $\sem{\formu_1 \lor \formu_2}{\calI, \alpha} := \F$.
  \ignore {
    \item $\sem{\formu_1 \limp \formu_2}{\calI, \alpha} := \T$
      if $\sem{\formu_1}{\calI, \alpha} = 0$ or $\sem{\formu_2}{\calI, \alpha} = \T$,
      else $\sem{\formu_1\limp \formu_2}{\calI, \alpha} := \F$.
  }
\end{itemize}

We say that an interpretation \emph{satisfies} a formula
$\phi$, if $\sem{\phi}{\calI, \alpha} = \T$.
If $\calI$ is fixed, the satisfiability of a formula depends only on
the assignment $\alpha$, and
$\phi$ is \emph{satisfiable} if there is an assignment $\alpha$ such that satisfies $\phi$.
Hence, we write $\sem{\phi}{\alpha}$ in place of $\sem{\phi}{\calI, \alpha}$ if
$\calI$ is clear from the context.

\begin{example}
  Let $t$ be
  $0.1  * \speciesE * \speciesS$. 
  This is a term over variables $\{\speciesE$, $\speciesS\}$, and the constant $0.1$.
  With the usual real interpretation $\calI_\setreals$, 
  and the assignment $\alpha(\speciesE) = 1$, $\alpha(\speciesS) = 2$, we get
  $\sem{t}{\calI_{\setR}, \alpha} = \{0.2\}$.
  Alternatively, the sign interpretation $\calI_{\setS}$ 
  maps constants and variables to the signs $\{\signUP, \signDO, \signNO\}$.
  Intuitively, $\signUP$ means that a value is positive, $\signDO$ denotes
  a negative value, and $\signNO$ a value that is exactly $0$.
  Using the sign assignment
  $\alpha(\speciesE) = 0$ and $\alpha(\speciesS) = \signUP$, we get
  $\sem{t}{\calI_\setreal, \alpha} = \{0\}$.
  The atom $t \EQ \speciesS * \speciesE * k$ is a formula that is
  satisfiable over the reals and the signs, for all assignments
  to $\speciesS$ and $\speciesE$.
\end{example}

\subsection{First-Order Boolean Networks with Non-deterministic Updates}

From a set of variables $V$
we build four sets of annotated variables
$\setSpecies = V $,
$\next\setSpecies = \{  \next{v} \mid v \in V  \} $,
$\dot\setSpecies = \{ \dot{v}\mid v \in V  \}$,
$\nextdot{\setSpecies} = \{\nextdot{v} \mid v \in V  \}$.
Furthermore,
$\setVars = \setSpecies \cup \next\setSpecies \cup \dot\setSpecies \cup \nextdot{\setSpecies}$.
The rationale here is that variables in $\setSpecies$ represent the species,
their value will model their amount of the species,
variables in $\next{\setSpecies}$ represent the amount of the species after a transition,
dotted variables in $\dot{\setSpecies}$ represent the derivative of the species,
and dotted next variables in $\nextdot{\setSpecies}$ represent the derivative after a transition.

\subsubsection{Syntax of FOBNN}
Formally, a FOBNN $\fobnn$ is a first-order formula $\phi$ over the signature $\SigmaArith$ and variables
in $\setVars$.
The formula must have the form $\exists v_1, \dots, v_n.\,\Omega$ where
$\{v_1, \dots, v_n\} = \dot\setSpecies \cup \nextdot{\setSpecies}$.
The variables in $\setSpecies \cup \setNSpecies$ remain free.
The formula $\Omega$ is a quantifier free conjunction of atoms
(see \Cref{ex:smallfobnn}).
Hence, the formula is $\Sigma_1$ form: 
the only quantifiers are existential quantifiers in the prefix.

\subsubsection{FOBNN State Transitions}
If we fix an interpretation $\calI$, then each model of $\fobnn$
is determined by an assignment $\alpha$ to the free variables $\setSpecies \cup \setNSpecies$. 
This represents a transition between two states $\state \to \next{\state}$
where $\state := \restr{\alpha}{\setSpecies}$
is the \emph{current state},
and $\next{\state} = \restr{\alpha}{\setNSpecies}$
is the \emph{next state}.
By enumerating the assignments that satisfy $\fobnn$ under $\calI$,
we build the \emph{base state transition graph} of the FOBNN $\fobnn$.
If we remove the existential quantifiers from $\fobnn$, the assignment $\alpha$
will assign values to all variables.
This represents a transition between two \emph{extended states} $\hat{\state} \to \next{\hat{\state}}$ where
$\hat{\state} := \restr{\alpha}{\setSpecies{\cup \setDSpec}}$
and $\next{\hat{\state}} = \restr{\alpha}{\setNspecies{\cup \setDNSpec}}$.
The set of such transitions is the \emph{extended state transition graph} of $\phi$.

Which interpretation should we use?
We discuss below that
we can use real numbers and the usual arithmetic operators
to model reaction networks.
However,
\citeauthor{niehrenAbstractSimulationReaction2022}~\cite{niehrenAbstractSimulationReaction2022}
use the abstract set of signs.
This allows us
to reason about the FOBNN efficiently, and
to obtain a boolean state transition graph.

In this case, FOBNNs are interpreted by a triple $(\setS, \calI, \alpha)$
where $\setS$ is the set of signs $\{\signUP, \signDO, \signNO\}$.
\begin{itemize}
    \item
          The interpretation of constants is their sign.
          For example, $\calI(-2.3) = \signDO$.
    \item
          The interpretation of an arithmetic operator $\circ \in \SigmaArith$
          is a subset of $\setS^3$, such that
          for all real numbers $x_1$, $x_2$, $x$ with signs $s_1$, $s_2$, $s$
          and $x = x_1 \circ x_2$
          then $(s_1, s_2, s) \in \calI(\circ)$.
          For example, $\{(\signUP, \signDO, \signUP), (\signUP, \signDO, \signDO), (\signUP, \signDO, \signNO)\}
              \subset \calI(+)$.
\end{itemize}

\subsubsection{FOBNNs from Reaction Network ODEs}
\label{sec:fobnnsfromrn}

FOBNNs are typically used to reason about dynamical systems.
In this case, they are built from the ODEs of a reaction network.

Let $\setSpecies$ be a set of species. 
The amount of chemicals in a solution is modeled as a 
\emph{concrete base state}
 $\sigma : \setSpecies \to \setRp$.
Base states are written in the standard notation adopted from chemistry:
$\sum_{\speciesX \in \setSpecies} \sigma(\speciesX) * \speciesX$. 
For example the state 
$\sigma(\speciesA) = 1.2, \sigma(\speciesB) = 3$ 
is written $1.2 * \speciesA + 3 * \speciesB$. 
Note that this is an arithmetic term
from $\setTermsArith$.
A reaction $(\rho, e, \pi)$ describes the transformation of 
a pool $\rho$ of reactants
into a pool $\pi$ of products, 
that happens at a speed $e$.
The elements $\rho$ and $\pi$ are concrete base states over $\setSpecies$.
Formally, a reaction is thus an element of 
$\setRp^{\setSpecies}\times \setTermsArith \times \setRp^{\setSpecies}$.

A \emph{reaction network}
is a set of reactions.
Any reaction network $\rn$ is
associated with an ordinary differential equation system $\odeR$.
ODEs can be represented as first-order formulas over the signature
$\setR \cup \arithmOps \cup \{\dot{\phantom{c}}\}$.
Those formulas are interpreted in the universe of
real valued functions $\setRp \to \setR$.
Constants are interpreted with constant functions, and the interpretation
of operators combine functions point-wise.
The interpretation of the dot-operator is such that $\calI(\dot{\phantom{c}})(f)$ is the
derivative of $f$ for any differentiable function, and undefined otherwise.
An ODE system consists of a conjunction of atomic formulas (equations) over this
signature with the dot-operator applied to variables
on the left-hand side,
and an arithmetic term on the right-hand side:
\[
  \odes_{\rn} \eqdef
  \bigland_{\speciesX\in\setSpecies}
  \dotX \EQ \sum_{(\rho,e,\pi) \in \rn} (\rho(\speciesX) - \pi(\speciesX)) * e
\]
where $\rho(\speciesX)$ and $\pi(\speciesX)$ are the constant symbols
representing the quantity of species $\speciesX$ as reactant and product respectively.

\begin{example}
  \label{example-Renz}
  \label{fig-Renz}

  The reaction network $\Renz$
  models a simple enzymatic process using
  three reactions over a set of four species
  $\setSpecies = \{\speciesS, \speciesE, \speciesES$, $\speciesP\}$:
  a molecule of a substrate $\speciesS$ binds reversibly an enzyme $\speciesE$
  to form a complex $\speciesC$ ($\ron$ and $\roff$).
  The complex $\speciesC$ can transform into the free enzyme $\speciesE$
  and two molecules of a product $\speciesP$:

{
    \centering
    $
      \Renz := \{
      \ron: \speciesS + \speciesE
      \xrightarrow{ \varkon * \speciesS * \speciesE }
      \speciesC ;
      \roff: \speciesC
      \xrightarrow{ \varkoff * \speciesC}
      \speciesS + \speciesE ;
      \rcat: \speciesC
      \xrightarrow{ \varkcat * \speciesC}
      \speciesE + 2 * \speciesP
      \}
    $
}

\noindent
  The kinetic expression for each reaction is an arithmetic term from $\setTerms_{\SigmaArith, \setSpecies}$. 
  It is indicated atop the arrow.
  These use the constants $\varkon$, $\varkoff$, $\varkcat$,
  that are in $\setRp$.
\noindent
  The ODE associated with the species $\speciesC$ of $\Renz$ corresponds to the
  following first-order atom:
  $
        {\dot{\varspecies{C}} }
    \EQ %
    \eon - \eoff -
    \ecat
    .
  $%
  \footnote{For the full system see \cref{fig-Renz-odes}.}
\end{example}

Solutions for ODE systems
assign functions to $\speciesX$ and $\dotX$
that compute values for any time $t$ in the solution interval.
If we read $\dotX$ as a dotted variable instead of the dot-operator applied
to $\speciesX$, 
we can understand the ODE as a continuous transition system
over the variables ${\setspec \cup \dot{\setspec}}$.  
This system operates on states $\hat{\sigma}_t$ along time $t$.
Those are the \emph{concrete extended states} of the FOBNN,
while
$\state_t := {\hat{\state}}_{t|\setspec}$ %
are \emph{concrete base states}.
While the derivative $\hat{\state}_t(\dotX)$ can be negative, 
in our setting all $\speciesX \in \setSpecies$
represent an amount and
$\state_t(\speciesX) \geq 0$.

FOBNNs interpreted over the real numbers implicitly express this transition between
two states of an ODE.  Let $\reactnet$ be a reaction network with
species in $\setSpecies$, and $\odeR$ the ODE of $\reactnet$.
Its FOBNN consists of the conjunction of

\begin{itemize}
    \item the formula $\odeR$ (conjunction of atoms over $\setTerms_{\SigmaArith, \setVars}$);
    \item the formula $\next{\odeR}$,
          where $\next{\odeR}$ is  a copy of $\odeR$,
          with each variable $v$ replaced by a fresh variable $\next{v}$;
    \item for each species $\speciesX \in \setSpecies$,
          the formula
          $\speciesX \geq 0 \land \nextX \geq 0 \land \nextX \EQ \speciesX + \dotX$
          that imposes the expected non-negativity of the species variables (derivatives can be negative) and uses
          the \emph{current value} $\speciesX$ and the \emph{current differential} $\dotX$
          to computes the value of $\speciesX$ in the \emph{next state}.
\end{itemize}

\noindent
Overall, an FOBNN formula thus has the form:
\[
    \fobnn_{\reactnet} := (\exists \dot{\speciesX}.\, \exists \dotnext{\speciesX}.\,)_{\speciesX \in \setSpecies}
    \left(
    \odeR
    \land
    \next{\odeR}
    \land
    \bigwedge_{\speciesX \in \setSpecies} \nextX \EQ \speciesX + \dotX
    \right).
\]

\begin{example}
  	\label{ex:smallfobnn}
    The following formula is a subformula of $\phiRenz$, the FOBNN associated
    with the reaction network $\Renz$.%
    \footnote{The full FOBNN is given in \cref{fig-Renz-fobnn}.}
    $$
    \begin{aligned}
                 \dot{\varspecies{C}} &\EQ \eon - \eoff-\ecat
     \\ 
     \land \quad \next{\dot{\varspecies{C}}}  &\EQ \nexteon - \nexteoff - \nextecat
     \\
     \land  \quad     \next{\varspecies{C}}   &\EQ {\varspecies{C}} +\dot{\varspecies{C}} %
    \end{aligned}
    $$
\end{example}

FOBNNs interpreted with signs induce a finite and
non-deterministic state transition system
whose state transition relation is given implicitly.
There are $ 2^{|\setspec|}$ base states,
and $ 3^{|\setspec|} \times 2^{|\setspec|}$ extended states.
One base state corresponds to several extended states.
The transition graph
represents the dynamics of the state transition system.
Since the base state transition graph uses assignments of booleans ($0$, $\signUP$) to species,
it is similar to the familiar state transition graphs of classic boolean networks~\cite{schwab_concepts_2020}.
A common experiment on state transition graphs is to search for
\emph{fixed points}: self-looping state from which we cannot escape.

\subsection{Additional Constraints}
\label{sec:additional-constraints:kinetics}

The transition graph generated by an FOBNN contains spurious transitions compared
to the sign abstraction of the real transition graph of the ODE.
\citeauthor{niehrenAbstractSimulationReaction2022}~\cite{niehrenAbstractSimulationReaction2022}
proposed to eliminate some spurious transitions from the graph
by using additional knowledge.

One approach is to add formulas to the FOBNN that are entailed by its models
over the real numbers.  Since such formulas are not necessarily entailed by the
models over the signs, they can eliminate some spurious solutions.
In some cases this can even make the abstraction precise~\cite{allart21}.

Another approach is to encode domain knowledge.
For example, consider
a reaction network 
with one reaction $\speciesA + \speciesB \xrightarrow{{{k * \speciesA * \speciesB}}}{} \speciesC$.
The speed of this reaction follows the mass action law:
the rate of the reaction is
directly proportional to the product of the value for the reactants.
A consequence of using this law
is that it is impossible
for a species that is present ($\speciesX = \signUP$)
to become absent ($\speciesX = \signNO$).
With the naive construction of FOBNN, 
the computed transition graph would have the edge
$\signUP\signUP 0 \tto \signUP\signUP\signUP$, 
$\signUP\signUP 0 \tto \signUP0\signUP$,
$\signUP\signUP 0 \tto 0\signUP\signUP$,
$\signUP\signUP 0 \tto 000$. 
However, the first edge alone is a tighter sound abstraction
of the actual dynamics of the ODEs over the reals.
By adding $\next\speciesX \geq \speciesX$
to the formula $\fobnn$ for all the species $\speciesX$
whose dynamics are governed by a mass action law kinetics,
one can make the FOBNN abstraction tighter, and still correct.
A similar constraint could be added for other kinetics that do not allow species to go back to 0,
such as the Michaelis-Menten kinetics. 
Another example of this is on the reaction network $\Renz$, %
where the kinetics of the three reactions follows the mass action law.
\Cref{Renz-stg-MA,Renz-stg-noMA} are the transition graphs
without and with external knowledge about the mass action constraints.
One can see the later graph is smaller.

\section{Encoding FOBNNs as Propositional Formulas}
\label{sec:encoding}

As we just saw, FOBNNs use first-order logic to describe state transitions.
The models of this formula define a relation between the state variables,
and the next-state variables.
However, the typical interpretation is the fixed finite domain of signs.
In this case,
\emph{satisfiability is decidable}
and formulas can efficiently be
translated into  \emph{equisatisfiable propositional logic formulas}.

Propositional formulas are simpler than first-order formulas
in that they do not use terms nor quantifiers.
The propositional formulas $\psi$ are constructed using the production rule
\begin{multline*}
    \psi  ::=  v
    \mid  \neg \psi
    \mid \psi_1 \land \psi_2
    \mid \psi_1 \lor  \psi_2
    \mid \psi_1 \land \psi_2
    \\
    \text{where }v \in \setVars\text{ and }\psi_1, \psi_2 \text{ are propositional formulas.}
\end{multline*}

\noindent
The truth value of a propositional formula is determined
by an assignment $\beta$ that maps variables
directly to truth values ($\F$ or $\T$).
The interpretation $\sem{\psi}{\beta}$ of
a propositional formula $\psi$ is defined just as for first-order
logic.
However, the atoms are propositional variables whose truth value
is determined by $\beta$.
If there is a $\beta$
such that $\sem{\psi}{\beta} = \T$, then we call $\beta$
a \emph{model} and say that $\psi$ is \emph{satisfiable}.

This section describes the encoding of an FOBNN formula $\fobnn$ into a
propositional formula $\pbnn$.
The translations conceptually consists of two steps:
(1) we flatten the input FOBNN formula; (2) we encode the flat formula in propositional logic.
Note that we present those two phases separately for clarity, but 
our implementation interleaves both phases.

Our translation of a FOBNN $\fobnn$ to a propositional formula $\pbnn$ is sounds.
That means that each model of $\pbnn$ can be translated to a state transition of $\fobnn$,
and that $\pbnn$ is unsatisfiable, if $\fobnn$ has unsatisfiable with respect to the
sign interpretation.

\subsection{Flat-Term Translations}
The terms used by an FOBNN formula $\fobnn$ may be not flat, meaning that
they contain nested function application.

Conversely, a term $t$ is a \emph{flat term} if it can be produced by the rule
\[
t ::= v \mid c \mid f(v_1, v_2)\text{ where }v, v_1, v_2 \in \setVars\text{ and }c \in \setreals.
\]

Furthermore, a first-order formula is \emph{flat} if all its atoms have the shape
$v \EQ t$ where $v$ is a variable, and $t$ is a flat term.

The first step of the propositional encoding procedure is to translate an FOBNN
$\fobnn$ to an equisatisfiable and flat first-order formula $\fobnn'$.

The flat form uses additional variables.
We assume we have a large enough of set $\mathcal{W}$ of fresh variables
such that $\setVars \cap \mathcal{W} = \emptyset$.
Recall that $\fobnn$ has the shape $\exists_{\speciesX\in\setSpecies \cup \setNSpecies} \speciesX.\,\Omega$ where $\Omega$
is a conjunction of atoms.  
Since the fresh variables are not relevant for the state transitions represented by an
FOBNN, the flat formula $\fobnn'$ will have the shape
$\exists_{\speciesX\in\setSpecies \cup \setNSpecies} \speciesX.\exists_{w \in \mathcal{W}} w.\,\Omega'$. 

The flattening procedure transforms nested terms into flat terms by introducing
fresh variables and auxiliary equations at each level of the term tree.
This means, to build $\Omega'$ we
\begin{enumerate}
  \item iterate over the atoms $t_1 \EQ t_2$ of $\Omega$,
  \item traverse the subterms of $t_1$ and $t_2$ bottom up, and
  	\begin{enumerate}
      \item for each $c \in \setreals$ encountered, add $w \EQ c$ to $\Omega'$ where $w\in\mathcal{W}$ is fresh,
      \item for each $f(t_1, t_2)$ encountered, add $w \EQ f(w_1, w_2)$ to $\Omega'$ where $w_1, w_2, w\in\mathcal{W}$, $w$ is
      	fresh, and $w_1$, $w_2$ are the fresh variables assigned to $t_1$ and $t_2$.
  	\end{enumerate}
\end{enumerate}

\begin{example}

The flat version of 
$\exists \dot \speciesC.\,
\dot \speciesC \EQ \varkon * \speciesS * \speciesE - \varkoff * \speciesC - \varkcat * \speciesC$
is
\begin{multline*}
\exists \dot \speciesC.\,\exists
w_1, w_2, w_3, w_4, w_5, w_6, w_7, w_8, w_9.\\
		  w_1   \EQ \varkon
\land w_2   \EQ w_1 * \speciesS
\land w_3   \EQ w_2 * \speciesE
\land w_4   \EQ \varkoff  
\land w_5   \EQ w_4 * \speciesC\\
\land w_6   \EQ w_3 - w_5
\land w_7   \EQ \varkcat
\land w_8   \EQ w_7 * \speciesC
\land w_9   \EQ w_6 - w_8
\land \dot \speciesC \EQ w_9.
\end{multline*}

\end{example}

\subsection{Encoding in Propositional Logic}

Once the FOBNN formula $\fobnn$ has been flattened to $\fobnn'$, we translate it
into a propositional formula $\pbnn$.
We assign two propositional variables $v^0$, $v^1$ to each
variable in $v \in \setVars \cup \mathcal{W}$.
These two proportional variables encode a sign. 
Given a propositional assignment $\beta$,
we define a translation function to signs:
\[
  \sign(\beta)(v^0, v^1) = \begin{cases*}
  	\signUP & if $\beta(v^0) = \T$ and $\beta(v^1) = \F$ \\ 
  	\signDO & if $\beta(v^0) = \F$ and $\beta(v^1) = \T$ \\
  	\signNO & if $\beta(v^0) = \F$ and $\beta(v^1) = \F$ \\ 
	\end{cases*}
\]

\noindent
Note that $\sign(\beta)(v^0, v^1)$ is undefined if
$\beta(v^0) = \beta(v^1) = \T$.
Hence, the encoding has to ensure such a $\beta$ is never a model of
$\pbnn$.

We can now state precisely when the encoding is sound.
Overall, the formulas $\fobnn$ and $\pbnn$ must be \emph{equisatisfiable}:
$\fobnn$ has a first-order model $\alpha$ 
\emph{if and only if} $\pbnn$ has a model $\beta$.  Furthermore, the models
must agree.  If $\beta$ is a model of $\pbnn$, then there
is a model $\alpha$ of $\fobnn'$ with $\sign(\beta)(v^0, v^1) = \alpha(v)$
for all free variables $v$ of $\fobnn'$.

Now assume, for example, that a model $\alpha$ is such that $\alpha(v) = \signUP$.
We can enforce that $\sign(\beta)(v^0, v^1) = \signUP$ by using
the formula $v^0 \land \neg v^1$.
We write $v = \signUP$ to denote this
formula.   Furthermore, $v = \signDO$ is $\neg v^0 \land v^1$, and
$v =\signNO$ is $\neg v^0 \land \neg v^1$

To encode a flat atom $v \EQ f(v_1, v_2)$ we have to ensure that
$(\sign(\beta)(v_1^0,v_1^1)$, $\sign(\beta)(v_2^0,v_2^1)$, $\sign(\beta)(v^0,v^1)) \in
\calI(f)$
for all
models $\beta$.
A naïve way to enforce this is the following encoding schema that enumerates
all elements of the relation:

\[
\bigvee_{(s_1,s_2,s) \in \calI(f)} v_1 = s_1 \land v_2 = s_2 \land v= s.
\]

Recall that $\fobnn'$ has the form
$\exists_{\speciesX\in\setSpecies \cup \setNSpecies} \speciesX.\exists_{w \in \mathcal{W}} w.\,\Omega'$ 
where $\Omega'$ is a \emph{conjunction} of flat atoms.
We can encode $\Omega'$ by forming the conjunction of the schema for each
individual atom.

Unfortunately, propositional logic does not allow us to encode the existential quantifiers,
and we have to omit them.
Consequently, each model $\alpha$ of $\fobnn$ corresponds to multiple
models $\beta$ of $\pbnn$.  We have to take this into account when we enumerate
models to compute transition graphs, as we describe in \Cref{sec:reasoning}.

\subsubsection{Practical Encoding}
While the encoding we described so far is sound and complete, it is not practical.
On the one hand, it introduces a conjunct for each triple $(s_1, s_2, s) \in \calI(f)$.
Those often can be combined and simplified.
On the other hand, the encoded formula is not in conjunctive normal form (CNF).
Most SAT solvers expect input problems as such a conjunction of disjunctions.
Hence, we use CNF formulas that are \emph{equivalent} to the schema that encodes atoms.

For example, when encoding an addition $v \EQ v_1 + v_2$, we know that a necessary condition for
$\sem{v_1 + v_2}{\alpha} = {\signUP}$
is that either
$\sem{v_1}{\alpha} = {\signUP}$ or $\sem{v_2}{\alpha} = {\signUP}$.
Hence, $v^0$ cannot be $\T$ if neither $v_1^0$ nor $v_2^0$ are $1$.
The dual holds for $\signDO$. This gives formula
\[
A := (v_1^0 \lor v_2^0 \lor \neg v^0) \land
     (v_1^1 \lor v_2^1 \lor \neg v^1).
\]

Furthermore, a sufficient condition for $\sem{v_1 + v_2}{\alpha}=\signUP$ is that one
of $v_1$, $v_2$ is $\signUP$, and the other one is $\signUP$ or $\signNO$.
Propositionally:
$v^0$ is $\T$ if $v_1^0 = \T$ and $v_2^1 = \F$,
or if $v_1^1 = \T$ and $v_2^0 = \F$.
This also holds for $\signDO$ with $v^1$, if
we flip $(.)^0$ and $(.)^1$.
Overall, we use the formula
\[
   (\neg{v_1}^0 \lor {v_2}^1 \lor v^0) \land
   (\neg{v_2}^0 \lor {v_1}^1 \lor v^0) \land
   (\neg{v_1}^1 \lor {v_2}^0 \lor v^1) \land
   (\neg{v_2}^1 \lor {v_1}^0 \lor v^1) \land A.
\]
Note that this formula is now a CNF formula. 
We use similar encodings for all operators.

\subsubsection{Sharing Variables is Unsound} To optimize the encoding, one might
consider reusing propositional variables for syntactically equal subterms.
However, this is unsound in the semantics of FOBNNs.  Consider the formula
$(\speciesX + \speciesY) * (\speciesX + \speciesY) \EQ -1$.  This formula is
satisfied by the assignment $\alpha(\speciesX) = \signUP$, $\alpha(\speciesY) = \signDO$,
because $\sem{\speciesX + \speciesY}{\alpha} = \setS$ and consequently
$\sem{(\speciesX + \speciesY) * (\speciesX + \speciesY)}{\alpha} = \setS$.
However, if the encodings of both occurrences of $\speciesX + \speciesY$ share
propositional variables, the sign of both sides of the product will be the same
for all models $\beta$, and no propositional model will interpret the
encoding of the product with $\signDO$.

\section{Incremental Reasoning About FOBNN Dynamics}
\label{sec:reasoning}

Modern SAT solvers support an \emph{incremental} interface~\cite{satrace15}
that allows users to modify SAT problem
while maintaining the solving state.
We can use this interface to explore the
dynamics of the FOBNN.

Two operations are possible:
(1) the addition of additional clauses;
(2) the use of extra literals as \emph{assumptions}.
Both operations modify the problem to solve,
but in the first case, the clauses are added permanently
while in the second case
the SAT solver will discard the assumptions after each query.

\subsubsection{Enumerating Transitions}
The most basic form of systematic network exploration is to list all
outgoing transitions of any state.
This can be implemented efficiently by using the incremental
interfaces by adding helper clauses.

After adding $\pbnn$ to the SAT solver as described in \Cref{sec:encoding},
any
model $\beta$ of $\pbnn$ corresponds to a transition of $\fobnn$.
Then, after obtaining a model $\beta$, we add clauses to the SAT solver
such that the formula is satisfiable by all transitions except
$\beta$.
The necessary clauses contain two literals that invert the assignment to
the variables $\speciesX \in \setSpecies \cup \next{\setSpecies}$:
add $\neg \speciesX^0   \lor \speciesX^1$  if $\sign(\beta)(\speciesX^0, \speciesX^1) = \signUP$,
add ${\speciesX}^0 \lor \neg \speciesX^1$  if $\sign(\beta)(\speciesX^0, \speciesX^1) = \signDO$,
and add ${\speciesX}^0 \lor     {\speciesX}^1$ if $\sign(\beta)(\speciesX^0, \speciesX^1) = \signNO$.
This enforces that the next model assigns a different value to at least one variable.
We can repeat this process until the formula is unsatisfiable.
This approach to model enumeration is standard.

To enforce that a transition starts at a certain
state we can use assumptions.
A state $\sigma$ maps the variables $\speciesX \in \setSpecies$ to signs.
The assumptions that enforce that every SAT model is a transitions starting
at $\sigma$ have to enforce that the model assigns the propositional values
corresponding to the signs.
The encoding is the same as for constants.
We assume exactly two literals for each
$\speciesX \in \setSpecies$:
${\speciesX}^0$ and $\neg{\speciesX}^1$ if $\sigma(\speciesX) = \signUP$,
$\neg{\speciesX}^0$ and ${\speciesX}^1$ if $\sigma(\speciesX) = \signDO$,
$\neg{\speciesX}^0$ and $\neg{\speciesX}^1$ if $\sigma(\speciesX) = \signNO$.

It is also possible to combine both approaches to enumerate, for example,
all outgoing transitions of a state, or to find fixed points.

\subsubsection{Fixed Points}
\label{sec:fixed-points}
The search for fixed points consists of two steps.
First, we find a loop, i.e., a transition where start and end state are the same.
Then, we check if the loop is a proper fixed point.

To find loops we add clauses that encode the constraints %
$\speciesX \EQ \next{\speciesX}$ for all $\speciesX \in \setSpecies$.
Now, each model is a loop.
To test whether a loop on state $\sigma$ is a fixed point we check
if there is a transition $\sigma \to \sigma'$ such that
there is $\speciesX \in \setSpecies$ with $\sigma(\speciesX) \neq \sigma'(\nextX)$.
To do this, we force the solver
to produce a transition starting in $\sigma$, as described above.
This query is \emph{unsatisfiable} if $\sigma$ is a fixed point.

However,  the constraints $\speciesX \EQ \next{\speciesX}$ are added permanently
to the SAT solver, and  we must be able to ``turn them off'' when
checking whether there is an outgoing transition.
To do so we create an additional propositional variable $v_{\mathrm{flag}}$ and
encode $\speciesX \EQ \next{\speciesX}$ as
$(\neg v_{\mathrm{flag}} \lor \neg\speciesX^0 \lor \next{\speciesX}^0) \land
(\neg v_{\mathrm{flag}} \lor \neg\next{\speciesX}^0 \lor \speciesX^0) \land
(\neg v_{\mathrm{flag}} \lor \neg\speciesX^1 \lor \next{\speciesX}^1) \land
(\neg v_{\mathrm{flag}} \lor \neg\next{\speciesX}^1 \lor \speciesX^1)$.

If $v_{\mathrm{flag}}$ is not assumed, the constraints can be satisfied
trivially by assigning $\F$ to $v_{\mathrm{flag}}$, and they have no further
effect on the model.
However, when $v_{\mathrm{flag}}$ is temporarily assumed during an SAT solver call,
then $\neg v_{\mathrm{flag}}$ is $\F$, and the loop constraints are active.

Overall, we search fixed points by first solving the SAT formula under the assumption
$v_{\mathrm{flag}}$ to obtain a candidate loop model $\beta$, and then checking whether
there is an outgoing edge from the state of $\beta$ by omitting the assumption $v_{\mathrm{flag}}$.
If this second SAT call is satisfiable, $\beta$
is not a fixed point.  In this case we exclude $\beta$ as described above,
and we look for another loop by assuming $v_{\mathrm{flag}}$ again.

\section{Evaluation}
\label{sec:evaluation}

\begin{figure}[t]
    \begin{subfigure}[h]{0.5\linewidth}
        \scalebox{0.45}{\input{fig/generated/cactus.pgf}}
        \caption{
            \label{fig-cactus}
            Solved models vs. runtime}
    \end{subfigure}
    \begin{subfigure}[h]{0.5\linewidth}
        \scalebox{0.45}{\input{fig/generated/time-vs-species.pgf}}
        \caption{
            \label{fig-time-vs-species}
            Number of species vs. runtime}
    \end{subfigure}
    \caption{
        Runtime (in ms) of FOBNN queries
        \AltTextCMSB{
            Left : cactus plot of solved models vs runtime
            Right : scatter plot of number of species vs runtime
        }
    }
\end{figure}

Our evaluation of the encoding is twofold.
First, in \Cref{sec:evaluation-practicality}
we used the encoding
to enumerate transitions and fixed points on several models.
The observed running time of these queries shows that
the encoding is practical.
We also discuss how our pipeline relates to a previous implementation that
uses constraint programming~\cite{niehrenAbstractSimulationReaction2022}.
Second, in \Cref{sec:evaluation-vsBiomodels},
we try to understand the FOBNN abstraction better by
comparing its state transition graphs
with those from the classic boolean semantics of reaction networks~\cite{fagesAbstractInterpretationTypes2008,calzone_biocham_2006}.
We show that our abstraction 
captures relevant properties of the dynamics of the reaction network.

\subsubsection{Implementation}
To perform large scale experiments, we implemented the propositional translation
procedure as part of a Python pipeline.
Our pipeline implements a robust parser for the CoreSBML~\cite{niehrenCoreSBMLIts2023}
language, an alternative to SBML.
It can encode reaction networks and comes with a formal definition of the
differential semantics
that simplifies the construction of the ODE system from the reaction network.
To interface with SAT solvers we use the Python library
PySAT~\cite{ignatievPySATPythonToolkit2018}.
We use its low-level interface that manipulates clauses as lists of integers. 
Hence, our implementation could be ported easily to other programming languages.
PySAT supports multiple SAT solvers.
We chose the state-of-the-art solver CaDiCal 1.0.3~\cite{BiereFazekasFleuryHeisinger-SAT-Competition-2020-solvers}
to perform all of our experiments.

\subsubsection{Dataset}
For our experiments, 
we use version 1.12 of the CoreSBML archive.\footnote{Available at \url{http://researchers.lille.inria.fr/niehren/Core-SBML/}}
It contains 541 models translated from the BioModels repository~\cite{malik-sheriffBioModels15Years2019}.
We discard models that use
noncontinuous elements
(events, or piece-wise functions), %
delays,
or local terms\footnote{The \texttt{apply} term.}.
In total, we are left with 198 models.
From them, 58 are used in \Cref{sec:evaluation-vsBiomodels}
as they explicitly use the concept of reaction to define the dynamics of the species,
as needed by the classic boolean semantics of reaction networks.

\subsection{Practicality of the Encoding}
\label{sec:evaluation-practicality}

\subsubsection{Methodology}

To assess the practicality of the encoding, we measured the runtime
of some simple queries.
First, we enumerated a fixed number of transitions ($1$, at most $50$, at most $100$) of
each model.
This type of query is the base for other analysis tasks,
such as visualizing full transition graphs, or enumerating the connections
of a particular state.
Second, we searched for a single fixed point.
This involves enumerating end eliminating loops.
Loops are common, bud fixed points are rare.
We enumerated at most 5000 loops. %
Runtimes were measured on a desktop computer with an
AMD Ryzen 9 7950X3D CPU and 64GB of RAM.

\subsubsection{Results.}

\Cref{fig-cactus} shows the number of solved instances in relation to the runtime.
We observed an expected behavior when enumerating transitions:
retrieving the first transitions takes longer than retrieving subsequent transitions.
Furthermore, finding fixed points is slower than simple enumeration.
\Cref{fig-time-vs-species} shows the solving time vs.\ the number of species.

The mean runtime for a single transition was 11ms, the maximum was 105ms.
The mean runtime for successful fixed-point searches was 34ms, the maximum was 902ms.
We could find fixed points in 85 models.

\begin{figure}[t]
    \begin{tikzpicture}[
            every node/.style={anchor=north west,inner sep=0pt},
        ]
        \node[draw=none,fill=none] at (0,0){\includegraphics[width=0.9\linewidth]{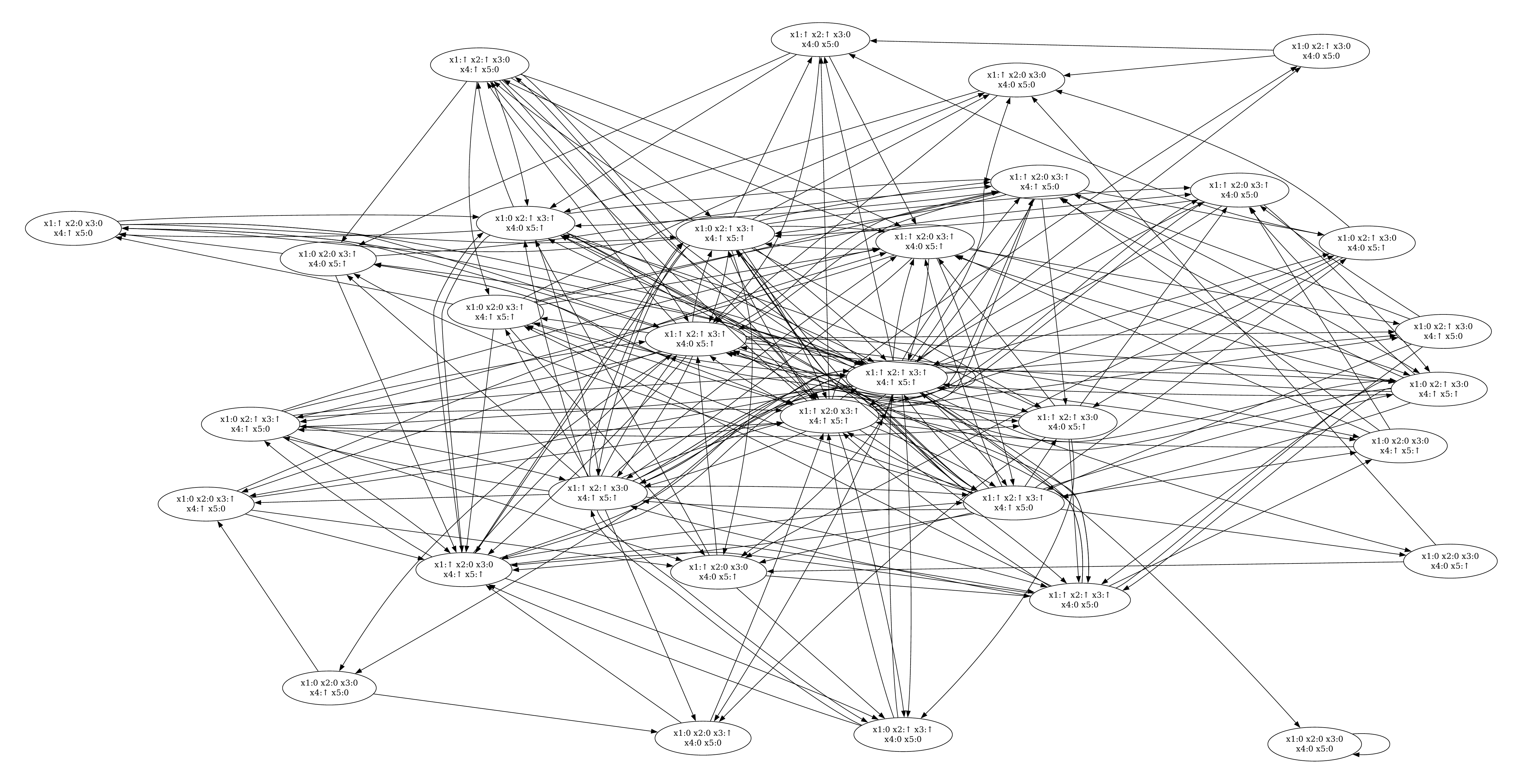}};
        \node[draw=none,fill=none] at (0,0){\includegraphics[width=8.5em]{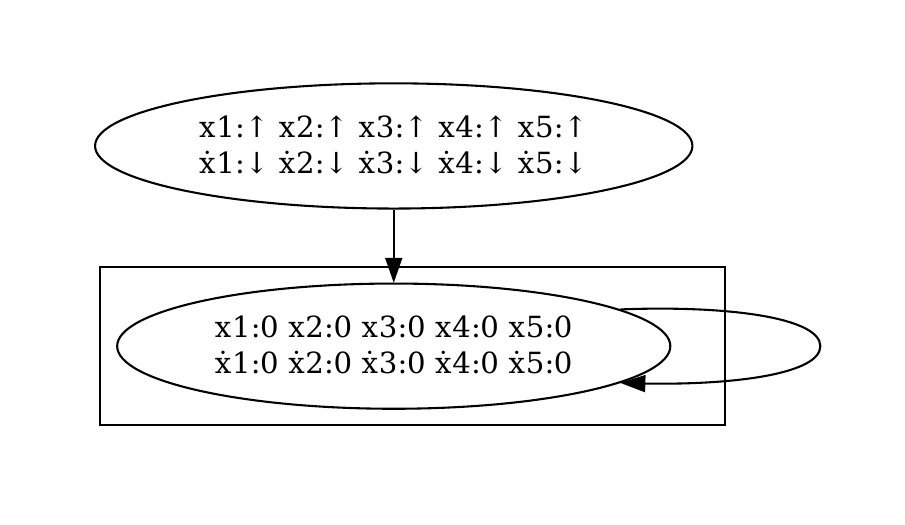}};
    \end{tikzpicture}
    \caption{
        \label{fig-full-state}
        Full state transition graph and fixed point of \emph{B197}.
        \AltTextCMSB{Full state transition graph and fixed point of \emph{B197}}
    }
\end{figure}

\Cref{fig-full-state} illustrates the full state transition graph
of the model \emph{B197}~\cite{bartholomeDatabasedMathematicalModeling2007}
This model contains five species $x_1$ to $x_5$.
Computing this graph took 20ms, and it
has $32$ nodes: one for each possible state.
Its density is $0.24$.
The \emph{density} of a graph is
the ratio between the number of edges in the graph
with number of edges in the complete graph.
There is a single fixed point where all species are $0$.
We confirmed numerically that this state corresponds to a stable steady-state
reached if the initial condition sets all the species to $0$.
The small graph in the top-left corner of \Cref{fig-full-state}
shows the extended states
associated with this fixed point.
Starting from the initial conditions provided in the published models
(all species at $0$ but for
$x_1 = 88$ nmol/ml), %
the system reaches a stable steady-state,
where all the species are present (are $\signUP$).
This state is in the center of the drawn graph, and
the FOBNN indeed finds a self-loop for this state.
Furthermore, this state stays a self-loop if we add constraints that
forces the variables $\dot{\speciesX}$ to $0$.
However, because of the undeterminism, this self-loop is not a fixed point.

\subsubsection{Reference Comparison}
As part of the proposal for FOBNNs,
\citeauthor{niehrenAbstractSimulationReaction2022}~\cite{niehrenAbstractSimulationReaction2022}
used constraint programming (CP) to compute transitions.
Like us, this implementation implements an FOBNN solver, and also
the translation from reaction networks to FOBNNs.
However, for various reasons both may
generate different FOBNNs from one reaction network.
Different additional constraints may be added, for instance, logical consequences over
reals (that are not consequences over the signs), but also
assumptions on forbidden states and state transitions.

There is currently neither a formally defined exchange format for FOBNNs
supported by both solvers, nor a conversion
between their formats. 
This makes it difficult to compare the
input FOBNNs for equality. Still, if computation terminates,
we can verify whether the computed transition graphs
are the same. To obtain a performance
indication nevertheless, we compared both solvers for FOBNNs of
few reaction networks. For the $\Renz$ example
reaction network, both solvers exhaustively enumerate the same
transition, and we can be sure that the underlying FOBNN formulas
are equivalent over signs. For the enumeration
the CP-based solver took 0.079s, and our solver took 0.003s (avg.\ of
10 runs). This example indicates that our solver is competitive with the
CP-solver.

\subsection{Comparing FOBNN with Classic Boolean Semantics}
\label{sec:evaluation-vsBiomodels}

\begin{figure}[t]
    \centering
    \includegraphics[width=0.80\textwidth]{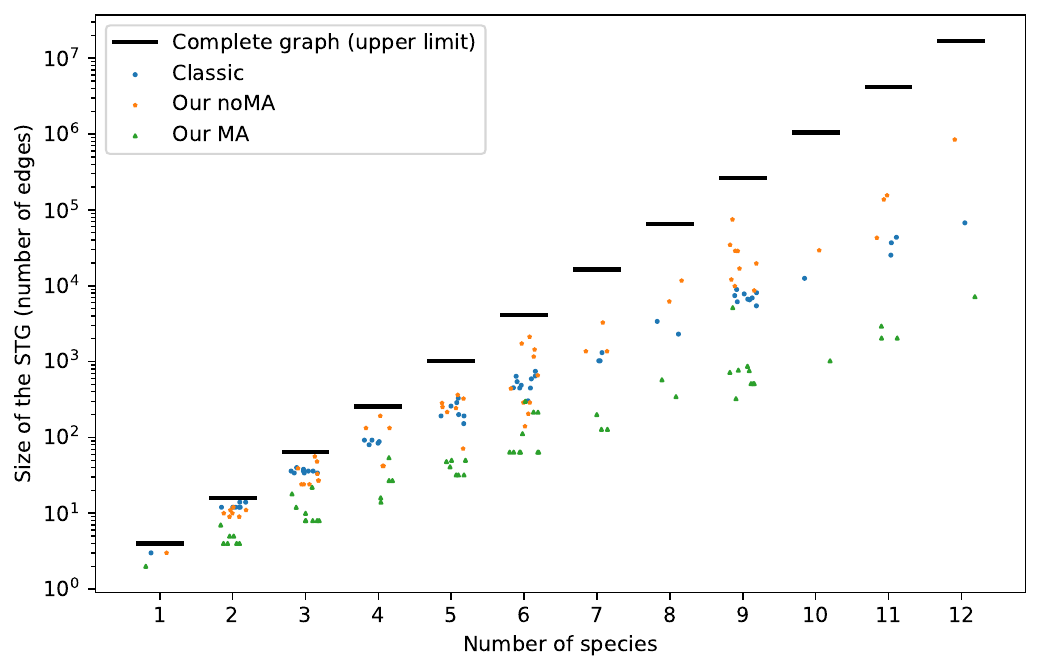}
    \caption{STG size with (\ourMA{}) and without (\ournoMA{}) mass action constraints \AltTextCMSB{Scatter plot of number of species vs. number of edges in the transition graphs}}
    \label{fig:compare-stg-edge-our-biocham}
\end{figure}

Both FOBNNs and the classic boolean abstraction of reaction networks~\cite{fagesAbstractInterpretationTypes2008} as implemented by Biocham~\cite{calzone_biocham_2006,}
produce boolean state transition graphs.

Given a reaction network $\reactnet$
over species $\setspecies$, 
and a state $\sigma$ where some species are present (are $\T$),
the classic boolean semantics selects (non-deterministically) at most one reaction 
$\reac \in \reactnet$ 
such that 
$\sigma(r) = \T$ for all reactants $r$ of $\reac$.
The products of $\react$ are guaranteed to be present in the next state, while its reactants may not persist. 
Uninvolved species keep their value.

A key problem when doing an abstraction is to refine it by eliminating spurious transitions, to better approximate the behavior of the original system. 
The classic boolean semantics of reaction networks provides a tight and correct abstraction of the stochastic semantics of reaction systems, but not their differential semantics.
However, the abstraction of the differential semantics of reaction networks provided by naive FOBNNs is correct but not tight. 
Yet, a key advantage of FOBNNs is that refinement can be naturally achieved by adding constraints. 

This subsection is dedicated to the empirical evaluation of
(1) the intuition that the transition graphs obtained by a naive FOBNN are similar in size (number of nodes and number of edges) than those computed by the classic boolean semantics, 
and (2) the impact of adding one constraint from domain knowledge (the mass action constraint, see \cref{sec:additional-constraints:kinetics}) on the graphs computed from FOBNNs.

We compute the full state transition graphs of all 58 models with up to 12 species.
The complete graphs is for models with 12 contain at most {16\,777\,216} edges.
Since Biocham does not support computing a full boolean transition graph,
we compute those using an ASP-program (\Cref{Appendix:booleanBiochamASP}).

The transition graphs 
of FOBNNs and the classic transition graphs (\classic{})
turn out to show substantial differences.
First, the \classic{}-graphs necessarily contain all base states,
while impossible states are omitted from FOBNN graphs.
These omitted states would falsify some constraints of the SAT encoding.
For example, in \emph{B275}, which has 4 species, 8 states do not appear in the FOBNN graph.
The reason is the species $\mathrm{RA}$ that appears in the denominator of the definition of the parameter $\rho$.
Hence, $\mathrm{RA}$ cannot be $0$, as enforced by our encoding.
However, the classic boolean semantics does not capture this.
Interestingly, COPASI does not report an error for
a numerical simulation starting from a state with $\mathrm{RA} = 0$.

\Cref{fig:compare-stg-edge-our-biocham} 
shows the number of edges in the graph relative to the number of species. 
We observe than the number of edges in the graphs computed by \classic{} and
\ournoMA{} is similar, but those computed with \ourMA{} have fewer edges.
As we discussed, %
using the mass action constraint may not be justified. 
Yet, adding such constraints
can be a useful first step to study an FOBNN, as they reduce the number of edges. %
Furthermore, in the future,
species that
admit the mass action constraint could be detected automatically.
Finally, none of the transition graph of the classic boolean semantics
had a fixed point.
However, the FOBNN transition graph of 24 models of the 58 models tested had at least one fixed point.

\section{Conclusion and Perspective}
\label{sec:ccl}

Overall, the novel FOBNN model is compelling because it soundly
abstracts the numerical Euler simulation.
We now also know that they can be handled in practice using
standard methods.
The encoded FOBNNs are
finite state transition systems with a \emph{propositional} representation of
the transition relation, and it is possible to use techniques
from SAT-based model checking~\cite{biere2018} on them.

This work suggests several future directions.
The immediate next step is to study classical dynamical properties, beyond fixed points.
In particular, SAT-based methods for detecting limit cycles
of classic
boolean networks~\cite{sohSATBasedMethodFinding2023} could be adapted to FOBNNs.
Moreover, it would be valuable to study how dynamical features
of the continuous system (such as steady state stability) can be anticipated
from the FOBNN abstraction. %
On the technical side,  it would be useful to have a well-specified interchange format for FOBBNs,
and a benchmark collection of FOBNNs, in particular those derived from CoreSBML models, with various additional constraints.
This would allow us to compare different FOBNN solvers on the same inputs systematically.

Another key perspective is to further position FOBNN within the hierarchy
of reaction network semantics.
Finally, we could also place
FOBNNs among the \emph{influence-based models} of biological systems.
In particular, they could be compared to classic boolean networks,
whose most-permissive semantics is known to capture the ODE dynamics~\cite{pauleve_reconciling_2020}.

\nanis{just to recall :I use todoFINAL to hide some todos (potentially) for final revision}

\begin{credits}
    \subsubsection{\ackname}
    The authors would like to thank the Computational Logic Center
    of the University of Iowa,
    as well as Joachim Niehren, Loïc Paulevé, Cesare Tinelli and Simon Vilmin for fruitful discussions.
    HJS was partially supported by a gift from AWS. 
    AV  acknowledges support from the french Agence Nationale pour la Recherche in the score of the project REBON (grand number: ANR-23-CE45-0008). 
    \subsubsection{\discintname}
    The authors have no competing interests.
\end{credits}

\printbibliography

\appendix
\renewcommand\thefigure{\thesection.\arabic{figure}}
\setcounter{figure}{0}

\newpage
\section{Appendix: $\Renz$}

\begin{figure}

    \begin{equation*}
        \begin{aligned}
            & \; {\dot \speciesS}
            \; \eq \;
            \, - \,  \eon %
            \, + \, \eoff %
            \\
            \land
            & \; {\dot \speciesE}
            \; \eq \;
            \, - \,  \eon %
            \, + \, \eoff %
            { \, + \,    {\varkcat * {\speciesES} }  }
            \\
            \land
            &\;  \dot{\speciesES}
            \; \eq \;
            {
                  \eon %
                 \, - \, \eoff %
                }
            {\, - \,    {\varkcat * {\speciesES} }}
            \\
            \land
            & \; {\dot \speciesP }
            \; \eq \;
            {2 * \varkcat * {\speciesES}}
            \\
        \end{aligned}
    \end{equation*}
    \caption{
        \label{fig-Renz-odes}
        The ODEs $\odes_{\Renz}$ associated with $\Renz$ (see \cref{fig-Renz}).
    }
\end{figure}

\begin{figure}[h]
$$
  \exists \dot{\varspecies{S}}.\, \exists\dotnext{\varspecies{S}}.\,
\exists \dot{\varspecies{E}}.\, \exists\dotnext{\varspecies{E}}.\,
\exists \dot{\varspecies{C}}.\, \exists\dotnext{\varspecies{C}}.\,
\exists \dot{\varspecies{P}}.\, \exists\dotnext{\varspecies{P}}.
$$

    \[\begin{aligned}
(
           & {\dot{\varspecies{S}}}
          \EQ~-\eon~+\eoff
          \\
          \land\quad&
            {\dot{\varspecies{E}}}
            \EQ~-\eon~+\eoff~+ \phantom {1 * }  \ecat
          \\
          \land\quad&
            {\dot{\varspecies{C}} }
            \EQ \phantom{~-} \eon~-\eoff~- \phantom {1 * } \ecat
          \\
          \land\quad&
           {\dot{\varspecies{P}}}
            \EQ \phantom{~-\eon~+\eoff~~+}   2 * \ecat
          \\ %
          \\
          \land\quad&
            {\next{\dot{\varspecies{S}}}}
            \EQ~-\next{\eon}+\next{\eoff}
          \\
          \land\quad&
            {\next{\dot{\varspecies{E}}}}
            \EQ~-\next{\eon}+\next{\eoff}~+ \phantom {1 * }  \next{\ecat}
          \\
          \land\quad&
            {\next{\dot{\varspecies{C}}}}
            \EQ \phantom{~-} \next{\eon}-\next{\eoff}~-\phantom {1 * }  \next{\ecat}
          \\
          \land\quad
           & {\next{\dot{\varspecies{P}}}}
            \EQ   \phantom{~-\next{\eon}~+\next{\eoff}~+}     2*   \next{\ecat}
          \\ %
           \\ %
          \land\quad
          & {\next{\varspecies{S}}}  \EQ {\varspecies{S}}+{\dot{\varspecies{S}}}
          \quad \land \quad \varspecies{S} \ge 0
          \\
          \land\quad
          &  {\next{\varspecies{E}}}  \EQ  {\varspecies{E}}+ {\dot{\varspecies{E}}}
          \quad \land \quad  \varspecies{E} \ge 0
          \\
          \land\quad
          &   {\next{\varspecies{C}}}  \EQ  {\varspecies{C}} +{\dot{\varspecies{C}}}
          \quad \land \quad \varspecies{C} \ge 0
          \\
          \quad\land\quad
          &   {\next{\varspecies{P}}}  \EQ {\varspecies{P}} +{\dot{\varspecies{P}}}
          \quad \land \quad  \varspecies{P} \ge 0
)
    \end{aligned}
    \]

    \caption{
        \label{fig-Renz-fobnn}
        The FOBNN $\phi_{\Renz}$
        built from \cref{fig-Renz-odes}
    }
\end{figure}

\begin{figure}[t]
    \includegraphics[width=\textwidth]{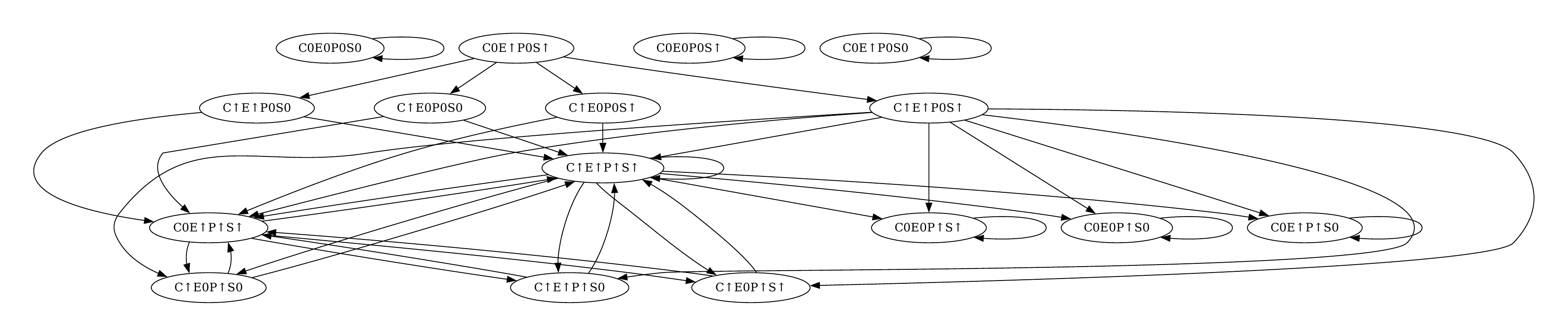}
    \caption{
    \label{fig-Renz-fobnn-stg-noMA}
    \label{Renz-stg-noMA}
    The transitions computed with the FOBNN for $\Renz$, without the mass action law kinetics constraint
    \AltTextCMSB{Transition graph for \Renz, without the mass action constraint}
    }
\end{figure}

\newpage
\section{Appendix: ASP encoding of the classic boolean semantics of reaction networks}


\begin{lstlisting}[style=myasp, caption={ASP encoding of the classic boolean semantics of reaction networks}, label={Appendix:booleanBiochamASP},
        basicstyle=\ttfamily\footnotesize,breaklines=true, %or \small or \footnotesize etc.
        %autodedent, 
         ]
% species are given directly by the input lp file.
%s(X) :- in(RID, X).
%s(X) :- out(RID, X).

% assert all the species have been declared :
:-  in(Rid, Sid); not s(Sid).
:- out(Rid, Sid); not s(Sid).

{present(0, S)} :- s(S).
state(0..1).

absent(STATE, X) :- not present(STATE, X); state(STATE); s(X) . 
-present(STATE, X) :- absent(STATE, X).

%%%%%%%%%%%%%%%%%%%%%%%%%%%%%
% If all the input of a reaction are present in STATE,
% then the reaction is firerable.
firerable(RID) :- present(0, X) : in(RID, X); s(X); r(RID).

% Among all the firerable reaction : pick one that is fired (usure asynchrony)
1{fired(RID) : firerable(RID)}1.
% if a reaction is firered, then all its outputs are present in STATE+1
present(1, Y) :- fired(RID); out(RID, Y).
% if a reaction is firered, then all its inputs are destroyable, but WE DONT KNOW if they will be present in STATE+1 -> use a choice rule.
destroyable(0, X) :- fired(RID); in(RID, X).
{present(1, X)} :- destroyable(0, X).

% If a species was present at state 0 and no reaction can destroy it, it is still there.
present(1, X) :- present(0, X); not destroyable(0, X).
% If a species was abscent at state 0 and no reaction can produce it, it is still not there.
% => nothing to do since the predicate present(1, X) will not be produced.
 
%%%%%%%%%%%%%%%%
#show present/2.
        \end{lstlisting}

\todoFINAL{use mint}

\end{document}